\documentclass[12pt]{article}
\usepackage{color}

\newcommand{\beq}{\begin{equation}}
\newcommand{\eeq}{\end{equation}}
\newcommand{\beqa}{\begin{eqnarray}}
\newcommand{\eeqa}{\end{eqnarray}}
\newcommand{\beqar}{\begin{eqnarray*}}
\newcommand{\eeqar}{\end{eqnarray*}}

\newcommand{\al}{\alpha}
\newcommand{\be}{\beta}

\def\spa          {\ \ \ }
\def\non          {\nonumber}

\def\s  {\sigma}
\def\spa          {\ \ \ }
\def\mand         {\spa\mbox{and}\spa}

\def\Tr           {\mbox{\rm Tr}\,}

\def\cd           {{\cdot}}
\def\ran          {\rangle}
\def\lan          {\langle}
\def\fsC    {C\!\!\!\!/\,}
\def\fsH    {H\!\!\!\!/\,}

\newcommand{\eps}{\epsilon}
\newcommand{\ga}{\gamma}

\newcommand{\inn}{\!\cdot\!}

\newcommand{\lam}{\lambda}

\newcommand{\z}{\zeta}

\newcommand{\labell}[1]{\label{#1}} 
\newcommand{\reef}[1]{(\ref{#1})}
\newcommand\prt{\partial}
\newcommand\veps{\varepsilon}

\newcommand\cL{{\cal L}}
\newcommand\cD{{\cal D}}

\newcommand\bz{\bar{z}}

\def\sst#1{{\scriptscriptstyle #1}}
\def\0{{\sst{(0)}}}
\def\1{{\sst{(1)}}}
\def\2{{\sst{(2)}}}
\def\3{{\sst{(3)}}}
\def\4{{\sst{(4)}}}
\def\5{{\sst{(5)}}}
\def\6{{\sst{(6)}}}
\def\7{{\sst{(7)}}}
\def\8{{\sst{(8)}}}

\begin{document}
\baselineskip 18pt%
\begin{titlepage}
\vspace*{1mm}%
\hfill

          
\vspace*{8mm}
\vspace*{8mm}%

\center{ {\bf \Large  
 	
On SYM theory and all order Bulk Singularity Structures of BPS Strings in type II theory

}}\vspace*{3mm} \centerline{{\Large {\bf  }}}
\vspace*{3mm}
\begin{center}
{Ehsan Hatefi  }

\vspace*{0.4cm}{\small  Institute for Theoretical Physics, TU Wien
\\
Wiedner Hauptstrasse 8-10/136, A-1040 Vienna, Austria\footnote{ehsan.hatefi@tuwien.ac.at,ehsan.hatefi@fuw.edu.pl }

\quad\\

Faculty of Physics, University of Warsaw, ul. Pasteura 5, 02-093 Warsaw, Poland
\quad\\

Mathematical Institute, Faculty of Mathematics,
Charles University, P-18675, CR}

\vskip.06in


\vspace*{.3cm}
\end{center}
\begin{center}{\bf Abstract}\end{center}
\begin{quote}

The complete forms of the S-matrix elements of a transverse scalar 
field, two world volume gauge fields, and a Potential  $C_{n-1}$ Ramond-Ramond (RR) form field are investigated.
In order to find an infinite number of $t,s, (t+s+u)$-channel bulk singularity structures of this particular mixed 
open-closed amplitude, we employ all the conformal field theory techniques to $<V_{C^{-2}}V_{\phi ^{0}} V_{A^{0}} V_{A^{0}}>$,
 exploring all the entire correlation functions and all order $\alpha'$ contact interactions to these  supersymmetric Yang-Mills (SYM) couplings.
Singularity and contact term comparisons with the other symmetric analysis, $<V_{C^{-1}}V_{\phi ^{0}} V_{A^{-1}} V_{A^{0}}>$ and 
$<V_{C^{-1}}V_{\phi ^{-1}} V_{A^{0}} V_{A^{0}}>$ are also  carried out in detail. Various couplings from pull-Back of 
branes, Myers terms and several generalized Bianchi identities should be taken into account to be able to 
reconstruct all order $\alpha'$ bulk singularities of type IIB (IIA) superstring theory.
Finally, we make a comment on how to derive without any ambiguity all order $\alpha'$ contact terms of this S-matrix which carry momentum of RR in transverse directions.
 
 \end{quote}
\end{titlepage}
\section{Introduction}
The non-perturbative and fundamental objects in type IIA and IIB superstring theories are called D$_p$-branes 
that have been first recognized in
\cite{Polchinski:1995mt}. Their role was highlighted in many different subjects, most specifically, by
theoretical high energy physics \cite{Witten:1995im}. As originally was pointed out, D$_p$-branes  include  $(p+1)$-dimensional 
world volume directions. They are hypersurfaces in all ten dimensional flat empty space that Neumann or Dirichlet boundary conditions can be applied to them. 

\vskip.08in

The effective action of multiple D$_p$-branes that embeds the commutator of transverse scalar fields was 
derived \cite{Myers:1999ps}. Making use of the direct mixed open-closed scattering amplitudes
with some reasonable field content, the generalization of this action to all
orders in $\alpha'$ has been done \cite{Hatefi:2012zh}.
Note that some anomalous couplings on D-branes as well as the phenomena  branes within branes have been discovered 
\cite{Green:1996dd} as well.
\vskip.08in
Based on doing direct Conformal Field Theory (CFT) techniques and scattering amplitude methods  several other couplings 
are also recently discovered in 
\cite{Hatefi:2015ora}. It was shown that  apart from three standard ways of exploring couplings in Effective Field Theory (EFT)
(pull-back of branes, Myers terms and Taylor expansions), one must introduce new EFT couplings to be able to reconstruct all the S-matrix of string amplitudes.

\vskip.1in

One can mention some of the applications for these  new couplings as follows. For example
in \cite{Hatefi:2012sy} we employed some of the new closed string interaction terms (Myers
terms) and constructed the leading $N^3$ entropy growth production from D0 quantum mechanics. We argued that if one would take into account some of Myers terms
(that have been taken from closed string RR couplings to scalar fields) then $N^3$ entropy behaviour of near
extremal M5 branes could have been computed from superYang-Mills couplings.  

\vskip.1in
Another example is related to AdS or dS solutions.  Based on Kaluza-Klein reduction formalism, in 
\cite{Hatefi:2012bp} we first dealt with ADM decomposition and then explicitly showed that the world volume theory can be described from the supergravity side. Basically we did apply this method to type IIB supergravity which was reduced on a 5 dimensional (5D) hyperboloidal space, and eventually discussed how one finds out either AdS or dS brane world
solutions by considering further reduction to 4D.

\vskip.1in
We also hope that some of these new EFT couplings may have some effects in the Large Volume Scenario (LVS) of IIB string compactifications. Indeed it is clarified in detail in \cite{deAlwis:2013gka}
 that if  we are in the domain that EFT holds then the LVS AdS minima are stable although they are non-supersymmetric.


\vskip.1in
Some important remarks for supersymmetrized version of this action were introduced \cite{Howe:2006rv}. Parts of the action that embeds symmetric trace of non-Abelian field content have been figured out.  
 In \cite{Leigh:1989jq} the action for single bosonic brane was suggested and eventually in \cite{Cederwall:1996pv}
the supersymmetric part of the action was derived. 
Notice that various remarks for the effective actions of brane anti brane systems, including
their bulk structures have also been revealed in \cite{Michel:2014lva}. It is worth pointing out  that most of RR
couplings are derived
through the mixed BPS open-closed  string amplitudes \cite{Kennedy:1999nn}.
To elaborate on all parts of the Chern-Simons, Wess-Zumino and DBI effective actions, we just refer to \cite{Hatefi:2012wj}.

\vskip.2in

The organization of the paper is as follows. In the next section we explicitly construct 
all the closed forms of the correlators for an asymmetric picture of the S-matrix
 $<V_{C^{-2}} V_{\phi^ {0}}  V_{A^{0}} V_{A^{0}}>$ where the superscripts indicate the related picture of strings.  Later on, for the concreteness  we demonstrate 
 the ultimate forms of S-matrix elements in two other symmetric pictures  
 $<V_{C^{-1}} V_{\phi^{-1}}  V_{A^{0}} V_{A^{0}}>$ and $<V_{C^{-1}} V_{\phi^{0}} V_{A^{-1}} V_{A^{0}}>$ as well. Note that, there is an alternative method \cite{Sen:2015hia} which can be applied to higher point functions.
 
\vskip.08in

Notations are described in the appendix and in references \cite{Hatefi:2015ora}  and \cite{Hatefi:2012wj}. The comparisons for all singularity structures and all order contact terms for both symmetric pictures have been performed in 
\cite{Hatefi:2015ora}. In this paper we first find out the complete
form of the S-matrix in asymmetric picture then carry out the comparisons for all singularity structures as well as 
 contact terms  with symmetric amplitude in detail.

\vskip.2in

   We explore various new couplings from $<V_{C^{-2}} V_{\phi^ {0}}  V_{A^{0}} V_{A^{0}}>$ amplitude and these couplings actually carry
 momentum of RR in transverse directions.  These new terms  include very explicitly  $p.\xi$ term in the ultimate  
 form of the amplitude. 
 We also  find out the presence of an infinite number of new $t,s,(t+s+u)$- channel bulk singularity structures which carry
momentum of RR in transverse directions, that is, $p^i$ terms just in an asymmetric picture of the amplitude.  These terms cannot be derived from the other pictures of amplitude as there are no winding modes ($w^i$ terms) in RR vertex operator in ten dimensions of spacetime. Finally we discover new contact terms in the asymmetric picture of the amplitude and these terms will be generated from an EFT point of view as well.
  \vskip.1in
  Further remarks are in order. All the correlation functions of $<e^{ip.x(z)} \partial x^i(x_1)> $ at asymmetric picture are non-zero thus one expects to generate new bulk singularity structures  from $<V_{C^{-2}} V_{\phi^ {0}}  V_{A^{0}} V_{A^{0}}>$ amplitude that cannot be appeared even in  $<V_{C^{-1}} V_{\phi^ {0}}  V_{A^{-1}} V_{A^{0}}>$ amplitude. Indeed by direct comparisons with  \cite{Hatefi:2015ora}, we show that just in an asymmetric amplitude, one is able to definitely construct all the bulk singularity structures of the S-matrix.
 
 Thus one needs to explore all the bulk singularities in the entire S-matrix elements that clearly involve the momentum of RR in 
 transverse directions. It is recently shown  in \cite{Hatefi:2015okf} that  even by knowing various
 generalized Bianchi identities for RR, one still is not able to discover all the bulk singularities of the simplest $<V_{C^{-1}} V_{\phi^ {0}}  V_{A^{-1}} V_{T^{0}}>$ amplitude.
 
   \vskip.1in
   
 In a recent attempt for the mixed closed RR-BPS  branes \cite{Hatefi:2016wof}  it is  shown that,  if one computes the amplitude (including an RR in the presence of transverse scalar field) in an asymmetric picture of RR and applies all the restricted Bianchi identities of C-field to the S-matrix elements then one is certainly able to precisely discover all the bulk singularity structures as well as all new contact terms of the string amplitude.

   \vskip.1in
 Therefore not only  asymmetric S-matrix has the potential to generate new structures
 for string couplings but also has the strong potential to  precisely fix all order $\alpha'$ corrections with 
 their exact  coefficients.

   \vskip.1in
   
 Hence, to be able to construct all the bulk singularity structures for the mixed RR and transverse scalar fields  one needs to find
 the complete form of the S-matrix analysis
 in asymmetric picture, explore restricted Bianchi identities and then start analyzing all the singularity structures. 
 It has clearly been observed that even the structures of some of the new string couplings can only be 
 figured out by direct S-matrix analysis not  by T-duality transformation.


\section{ The $<V_{C^{-2}} V_{\phi^ {0}}  V_{A^{0}} V_{A^{0}}>$ S-matrix}

Let us make some general observation about amplitudes at world-sheet level.
One might wonder about the deep relation between open and closed strings. This relationship has been widely understood once we try to match up all the string singularities  with their effective field theory parts. Based on various BPS and non-BPS five point functions 
in  \cite{Hatefi:2012rx}, we have 
made a universal conjecture that explores
  all order $\alpha'$  corrections to SYM couplings. This conjecture has come out of  exact reconstruction of 
all infinite  $(t+s+u)$-channel tachyonic or massless singularities, and having  an RR 
 coupling with a tachyon or scalar (gauge) field plays the major role in coinciding all the singularity structures 
 of both string and field theory.

\vskip.1in

Note that due to the appearance of the momentum of RR in transverse directions and the fact that winding modes are not inserted  in vertex of RR, one cannot apply
  T-duality transformation to S-matrix of $<V_{C}V_{A} V_AV_A>$ to get to complete form of $<V_{C}V_{\phi} V_AV_A>$ 
  S-matrix elements ( further explanations are given in \cite{Hatefi:2012ve},\cite{Park:2008sg}).
\vskip.08in

 Our computations are consist of the following vertex operators
  \beqa
V_{\phi}^{(0)}(x_1) &=&  \int dx_{1}\xi_{1i}\bigg(\partial
X^i(x_1)+\alpha' ik\cd\psi\psi^i(x_1)\bigg)e^{\alpha' ik\cd X(x_1)},
\nonumber\\
V_{\phi}^{(-1)}(y) &=& \int dy \xi.\psi(y) e^{-\phi(y)} e^{\alpha' ik\cd X(y)},
\nonumber\\
V_{A}^{(0)}(x_2) &=&  \int dx_{2}\xi_{2a}\bigg(\partial
X^a(x_2)+ \alpha'iq\cd\psi\psi^a(x_2)\bigg)e^{ \alpha' iq\cd X(x_2)},
\label{BPS12}\\
V_{C}^{(-\frac{3}{2},-\frac{1}{2})}(z,\bar{z})&=&(P_{-}\fsC_{(n-1)}M_p)^{\al\be} \int dz \int d\bar z e^{-3\phi(z)/2}
S_{\al}(z)e^{i\frac{\alpha'}{2}p\cd X(z)}e^{-\phi(\bar{z})/2} S_{\be}(\bar{z})
e^{i\frac{\alpha'}{2}p\cd D \cd X(\bar{z})}.
\nonumber
\eeqa
\vskip.1in

We also use kinematical relations as
\beqa
  k^2=q^2=p^2=0,  \quad q.\xi_{1}=k.\xi_1=q.\xi_2=0.
\nonumber\eeqa
where the RR vertex in asymmetric picture has been found in \cite{Bianchi:1991eu} and \cite{Liu:2001qa}.  One might also be interested in considering some of
the direct results of string amplitude computations \cite{Barreiro:2013dpa}. 
 
 \vskip.1in
 
The S-matrix elements of a  transverse scalar field, two world volume gauge fields and  one closed string C-field at disk level  
can be explored by  evaluating all the correlators  $<V_{C}^{(-\frac{3}{2},-\frac{1}{2})}(z,\bar{z}) V_{\phi}^{(0)}{(x_{1})}
V_{A}^{(0)}{(x_{2})}V_{A}^{(0)}{(x_{3})}>$ where all three open strings are inserted in zero picture and
the closed string RR is written down in terms of its potential.
\vskip.08in
First of all we try to explore the whole S-matrix, and to simplify the analysis we divide it out to various correlators.

\vskip 0.1in

To investigate  all  fermionic correlation functions, including a number of currents and also fermion fields
in different positions, 
one needs to employ the so called  generalized Wick-like rule which has been emphasized in \cite{Hatefi:2013yxa}.
Let us find  the compact form of the S-matrix elements \footnote{In type II we set $\alpha'=2$} as follows
\beqa
{\cal A}^{C^{-2}\phi ^{0}A^{0}A^{0}}&\sim&\Tr(\lam_1\lam_2\lam_3)\int dx_{1}dx_{2} dx_{3}dx_{4}dx_{5}(P_{-}\fsC_{(n-1)}M_p)^{\al\be}I\xi_{1i}\xi_{2a}\xi_{3b}x_{45}^{-3/4}\nonumber\\&&\times
\bigg((x_{45}^{-5/4} C^{-1}_{\alpha\beta})\bigg[ a^i_1a^a_2 a^b_3 -\eta^{ab} x_{23}^{-2} a^i_1\bigg]+i\alpha'  k_{2c} a^{i}_{1} a^{b}_{3} a^{ac}_{2}\nonumber\\&&
+i\alpha'  k_{1e} a^{ie}_{3} (-\eta^{ab} x_{23}^{-2}+a^a_2 a^b_3)-{\alpha'}^{2} k_{1e} k_{2c} a^b_3 a^{acie}_{4}+i\alpha'  k_{3d} a^i_1 a^a_2 a^{bd}_{5}\nonumber\\&&
-{\alpha'}^2 k_{3d} k_{2c} a^i_1 a^{bdac}_{6}-{\alpha'}^2 k_{3d} k_{1e} a^a_2  a^{bdie}_{7}-i{\alpha'}^3 k_{1e}k_{2c}k_{3d} a_8^{bdacie}\bigg)
\labell{amp3q333},\eeqa

where
$x_{ij}=x_i-x_j$, and also
\beqa
I&=&|x_{12}|^{\alpha'^2 k_1.k_2}|x_{13}|^{\alpha'^2 k_1.k_3}|x_{14}x_{15}|^{\frac{\alpha'^2}{2} k_1.p}|x_{23}|^{\alpha'^2 k_2.k_3}|
x_{24}x_{25}|^{\frac{\alpha'^2}{2} k_2.p}
|x_{34}x_{35}|^{\frac{\alpha'^2}{2} k_3.p}|x_{45}|^{\frac{\alpha'^2}{4}p.D.p},\nonumber\\
a^i_1&=&ip^{i}\bigg(\frac{x_{54}}{x_{14}x_{15}}\bigg)\nonumber\\
a^a_2&=&ik_1^{a}\bigg(\frac{x_{14}}{x_{24}x_{12}}+\frac{x_{15}}{x_{25}x_{12}}\bigg)+ik_3^{a}\bigg(\frac{x_{43}}{x_{24}x_{23}}+\frac{x_{53}}{x_{25}x_{23}}\bigg),\nonumber\\
a^b_3&=&ik_1^{b}\bigg(\frac{x_{14}}{x_{34}x_{13}}+\frac{x_{15}}{x_{35}x_{13}}\bigg)+ik_2^{b}\bigg(\frac{x_{24}}{x_{34}x_{23}}+\frac{x_{25}}{x_{35}x_{23}}\bigg),\nonumber\\
a^{ac}_{2}&=&2^{-1}x_{45}^{-1/4}(x_{24}x_{25})^{-1} (\Gamma^{ac}C^{-1})_{\alpha\beta} ,\nonumber\\
a^{ie}_{3}&=&2^{-1}x_{45}^{-1/4}(x_{14}x_{15})^{-1}(\Gamma^{ie}C^{-1})_{\alpha\beta} ,\nonumber\\
a^{acie}_{4}&=&2^{-2}x_{45}^{3/4}(x_{14}x_{15}x_{24}x_{25})^{-1}\bigg\{(\Gamma^{acie}C^{-1})_{\alpha\beta}+\alpha' m_1\frac{Re[x_{14}x_{25}]}{x_{12}x_{45}}\bigg\}
,\nonumber\\
a^{bd}_5&=&2^{-1}x_{45}^{-1/4}(x_{34}x_{35})^{-1} (\Gamma^{bd}C^{-1})_{\alpha\beta} ,\nonumber\\
a^{bdac}_6&=&2^{-2}x_{45}^{3/4}(x_{34}x_{35}x_{24}x_{25})^{-1}\bigg\{(\Gamma^{bdac}C^{-1})_{\alpha\beta}+\alpha' m_2\frac{Re[x_{24}x_{35}]}{x_{23}x_{45}}+\alpha'^2 m_3\bigg(\frac{Re[x_{24}x_{35}]}{x_{23}x_{45}}\bigg)^{2}\bigg\},\nonumber\\
a^{bdie}_{7}&=&2^{-2}x_{45}^{3/4}(x_{34}x_{35}x_{14}x_{15})^{-1}\bigg\{(\Gamma^{bdie}C^{-1})_{\alpha\beta}+\alpha' m_4\frac{Re[x_{14}x_{35}]}{x_{13}x_{45}}\bigg\}
\nonumber\eeqa
The  following expressions are introduced for the above correlators.
\beqa
m_1&=&\bigg(\eta^{ec}(\Gamma^{ai}C^{-1})_{\alpha\beta}
-\eta^{ea}(\Gamma^{ci}C^{-1})_{\alpha\beta}\bigg),\nonumber\\
m_2&=&\bigg(\eta^{cd}(\Gamma^{ba}C^{-1})_{\alpha\beta}
-\eta^{cb}(\Gamma^{da}C^{-1})_{\alpha\beta}-\eta^{ad}(\Gamma^{bc}C^{-1})_{\alpha\beta}
+\eta^{ab}(\Gamma^{dc}C^{-1})_{\alpha\beta}\bigg),\nonumber\\
m_3&=&(C^{-1})_{\alpha\beta}\bigg(-\eta^{ab}\eta^{cd}+\eta^{ad}\eta^{cb}\bigg),\nonumber\\
m_4&=&\bigg(\eta^{ed}(\Gamma^{bi}C^{-1})_{\alpha\beta}-\eta^{eb}(\Gamma^{di}C^{-1})_{\alpha\beta}\bigg).\eeqa
One also needs to elaborate on the last correlation function,  
$a_8^{bdacie}$ that has several terms as follows
\beqa
a_8^{bdacie}&=&<:S_{\al}(x_4):S_{\be}(x_5)::\psi^e\psi^i(x_1):\psi^c\psi^a(x_2):\psi^d\psi^b(x_3)>\label{33312}\eeqa
so that further ingredients are needed
\beqa
a_8^{bdacie}&=&
\bigg\{(\Gamma^{bdacie}C^{-1})_{{\alpha\beta}}+\alpha' m_5\frac{Re[x_{14}x_{25}]}{x_{12}x_{45}}+\alpha' m_6\frac{Re[x_{14}x_{35}]}{x_{13}x_{45}}+\alpha' m_7\frac{Re[x_{24}x_{35}]}{x_{23}x_{45}}
\nonumber\\&&+\alpha'^2 m_{8}\bigg(\frac{Re[x_{14}x_{25}]}{x_{12}x_{45}}\bigg)\bigg(\frac{Re[x_{24}x_{35}]}{x_{23}x_{45}}\bigg)
+\alpha'^2 m_9 \bigg(\frac{Re[x_{14}x_{35}]}{x_{13}x_{45}}\bigg)\bigg(\frac{Re[x_{24}x_{35}]}{x_{23}x_{45}}\bigg)
\nonumber\\&&+\alpha'^2 m_{10}\bigg(\frac{Re[x_{24}x_{35}]}{x_{23}x_{45}}\bigg)^{2}\bigg\}
\label{hh333}2^{-3}x_{45}^{7/4}(x_{14}x_{15}x_{24}x_{25}x_{34}x_{35})^{-1}\nonumber
\eeqa
where
\beqa
m_5&=&\bigg(\eta^{ec}(\Gamma^{bdai}C^{-1})_{\alpha\beta}
-\eta^{ea}(\Gamma^{bdci}C^{-1})_{\alpha\beta}\bigg),\nonumber\\
m_6&=&\bigg(\eta^{ed}(\Gamma^{baci}C^{-1})_{\alpha\beta}
-\eta^{eb}(\Gamma^{daci}C^{-1})_{\alpha\beta}\bigg),\nonumber\\
m_7&=&\bigg(\eta^{cd}(\Gamma^{baie}C^{-1})_{\alpha\beta}
-\eta^{cb}(\Gamma^{daie}C^{-1})_{\alpha\beta}
-\eta^{ad}(\Gamma^{bcie}C^{-1})_{\alpha\beta}
+\eta^{ab}(\Gamma^{dcie}C^{-1})_{\alpha\beta}\bigg),\nonumber\\
m_8&=&\bigg(-\eta^{ad}\eta^{ec}(\Gamma^{bi}C^{-1})_{\alpha\beta}
+\eta^{ab}\eta^{ec}(\Gamma^{di}C^{-1})_{\alpha\beta}
+\eta^{cd}\eta^{ea}(\Gamma^{bi}C^{-1})_{\alpha\beta}
-\eta^{cb}\eta^{ea}(\Gamma^{di}C^{-1})_{\alpha\beta}\bigg),\nonumber\\
m_9&=&\bigg(\eta^{ed}\eta^{cb}(\Gamma^{ai}C^{-1})_{\alpha\beta}-\eta^{ab}\eta^{ed}(\Gamma^{ci}C^{-1})_{\alpha\beta}
-\eta^{cd}\eta^{eb}(\Gamma^{ai}C^{-1})_{\alpha\beta}+\eta^{ad}\eta^{eb}(\Gamma^{ci}C^{-1})_{\alpha\beta}\bigg),\nonumber\\
m_{10}&=&\bigg(-\eta^{cd}\eta^{ab}(\Gamma^{ie}C^{-1})_{\alpha\beta}+\eta^{cb}\eta^{ad}(\Gamma^{ie}C^{-1})_{\alpha\beta}\bigg)\nonumber
\eeqa
Substituting the above correlators in the closed form of the S-matrix  and simplifying further, one finds 
that the amplitude is SL(2,R) invariant. The volume of conformal killing group has been  removed by fixing three position of 
open strings at  zero, one  and infinity, that is,  $ x_{1}=0 ,x_{2}=1,x_{3}\rightarrow \infty$.

One may find the evaluation of all the integrals on the location of closed string RR on upper half plane. 
The integrals can be computed only in terms of  Gamma functions  where one deals with the following sort of integrations on upper half plane 
\beqa
 \int d^2 \!z |1-z|^{a} |z|^{b} (z - \bar{z})^{c}
(z + \bar{z})^{d}\nonumber
\eeqa
where   $a,b,c$ are written just in terms of Mandelstam variables. For $ d= 0,1$ and $ d=2$ the
results are obtained in  \cite{Fotopoulos:2001pt} and \cite{Hatefi:2012wj} appropriately. We introduce the following definitions for all three Mandelstam variables as
\beqa
s&=&\frac{-\alpha'}{2}(k_1+k_3)^2,\quad t=\frac{-\alpha'}{2}(k_1+k_2)^2,\quad u=\frac{-\alpha'}{2}(k_2+k_3)^2
\nonumber\eeqa
Having done all the integrations properly, one would finally read off  all the elements of the string amplitude in asymmetric picture as follows
\beqa {\cal A}^{C^{-2}\phi^{0} A^{0} A^{0} }&=&{\cal A}_{1}+{\cal A}_{2}+{\cal A}_{3}+{\cal A}_{41}+{\cal A}_{42}+{\cal A}_{43}+{\cal A}_{5}\nonumber\\&&{\cal A}_{61}+{\cal A}_{62}
+{\cal A}_{63}+{\cal A}_{71}+{\cal A}_{72}+{\cal A}_{81}+{\cal A}_{82}\nonumber\\&&{\cal A}_{83}+{\cal A}_{84}+{\cal A}_{85}+{\cal A}_{86}+{\cal A}_{87}+{\cal A}_{88}
\labell{48ii}\eeqa
where
\beqa
{\cal A}_{1}&\!\!\!\sim\!\!\!&i p.\xi_1 \Tr(P_{-}\fsC_{(n-1)}M_p) L_1
\bigg[\frac{(-s-u)(-t-u)}{(-u-\frac{1}{2}) } \xi_{3}.\xi_{2} +4u k_1.\xi_{3} k_1.\xi_2\nonumber\\&& -4t k_3.\xi_{2} k_1.\xi_3 -4s  k_2.\xi_{3} k_1.\xi_2 
+\frac{2(t+s+u-2st)}{(-u-\frac{1}{2}) }  k_2.\xi_{3} k_3.\xi_2 \bigg],
\nonumber\\
{\cal A}_{2}&\sim&i k_{2c}\xi_{2a}p.\xi_1\Tr(P_{-}\fsC_{(n-1)}M_p \Gamma^{ac}) L_2
\bigg\{2ut k_1.\xi_{3} -2st k_2.\xi_{3}\bigg\}\nonumber\\
{\cal A}_{3}&\sim&i k_{1e}\xi_{1i}\Tr(P_{-}\fsC_{(n-1)}M_p \Gamma^{ie}) L_1
\bigg[-\frac{(-s-u)(-t-u)}{(-u-\frac{1}{2}) } \xi_{3}.\xi_{2} -4u k_1.\xi_{3} k_1.\xi_2\nonumber\\&& +4t k_3.\xi_{2} k_1.\xi_3 +4s  k_2.\xi_{3} k_1.\xi_2 
-\frac{2(t+s+u-2st)}{(-u-\frac{1}{2}) }  k_2.\xi_{3} k_3.\xi_2 \bigg]\nonumber\\
{\cal A}_{41}&\sim&i\Tr(P_{-}\fsC_{(n-1)}M_p \Gamma^{acie})\xi_{2a}\xi_{1i} k_{2c}k_{1e} L_2 \bigg(-2 ut k_1.\xi_3 +2 ts k_2.\xi_3\bigg)
\nonumber\\
{\cal A}_{42}&\sim& i \xi_{1i}  \xi_{2a} \Tr(P_{-}\fsC_{(n-1)}M_p \Gamma^{ai}) L_1\bigg\{-2ts k_2.\xi_{3}+2tu k_1.\xi_{3}  \bigg\}
\nonumber\\
{\cal A}_{43}&\sim& i \xi_{1i}  k_{2c} \Tr(P_{-}\fsC_{(n-1)}M_p \Gamma^{ci}) L_1\bigg\{-4s k_1.\xi_{2} k_2.\xi_{3}+4u k_1.\xi_{2} k_1.\xi_{3}  \bigg\}
\nonumber\\
{\cal A}_{5}&\sim&i p.\xi_1 k_{3d}\xi_{3b}\Tr(P_{-}\fsC_{(n-1)}M_p \Gamma^{bd}) L_2
\bigg\{-2 su k_1.\xi_{2}+2ts k_3.\xi_2\bigg\}\nonumber\\
{\cal A}_{61}&\sim&-4i\Tr(P_{-}\fsC_{(n-1)}M_p \Gamma^{dbac})\xi_{2a}\xi_{3b} p.\xi_1k_{3d}k_{2c}L_1 (-t-s-u)\nonumber\\
{\cal A}_{62}&\sim& -ip.\xi_1\Tr(P_{-}\fsC_{(n-1)}M_p \Gamma^{ba})  ts L_2\bigg\{-u\xi_{2a}\xi_{3b}-2k_2.\xi_3 k_{3b} \xi_{2a} -2k_3.\xi_2 k_{2a} \xi_{3b} 
+2\xi_{2}.\xi_{3} k_{2a}k_{3b}\bigg\}
\nonumber\\
{\cal A}_{63}&\sim& -ip.\xi_1  \Tr(P_{-}\fsC_{(n-1)}M_p )L_1  (u\xi_2.\xi_3+2k_3.\xi_2 k_2.\xi_3) \bigg(\frac{-2st+t+s+u}{(-u-\frac{1}{2})}\bigg)
\nonumber\\
{\cal A}_{71}&\sim&i \Tr(P_{-}\fsC_{(n-1)}M_p \Gamma^{bdie})\xi_{1i}\xi_{3b}  k_{3d}k_{1e} L_2 (-2ts k_3.\xi_2+2us k_1.\xi_2)\nonumber\\
{\cal A}_{72}&\sim& i \xi_{1i} L_1 \Tr(P_{-}\fsC_{(n-1)}M_p \Gamma^{bi})  \bigg\{2us\xi_{3b} k_1.\xi_{2}-2ts k_3.\xi_2 \xi_{3b} +4u k_1.\xi_{2}  k_1.\xi_{3}k_{3b}
-4t k_{3b} k_3.\xi_{2} k_1.\xi_{3}\bigg\}
\nonumber\\
{\cal A}_{81}&\sim& 4i \Tr(P_{-}\fsC_{(n-1)}M_p \Gamma^{bdacie})\xi_{1i}\xi_{2a}\xi_{3b}k_{1e} k_{2c}k_{3d} L_1 (-t-s-u)\nonumber\\
{\cal A}_{82}&\sim& -i su\Tr(P_{-}\fsC_{(n-1)}M_p \Gamma^{bdai})\xi_{1i}\xi_{3b} k_{3d} L_2\bigg\{t \xi_{2a}+2k_1.\xi_{2} k_{2a}\bigg\}
\nonumber\\
{\cal A}_{83}&\sim&  -i tu\Tr(P_{-}\fsC_{(n-1)}M_p \Gamma^{baci})\xi_{1i}\xi_{2a}k_{2c} L_2\bigg\{-s \xi_{3b}-2k_1.\xi_{3} k_{3b}\bigg\}
\nonumber\\
{\cal A}_{84}&\sim& ist \Tr(P_{-}\fsC_{(n-1)}M_p \Gamma^{baie}) k_{1e}\xi_{1i}\xi_{2a} L_2\bigg\{-u\xi_{3b}-2 k_{2}.\xi_{3} k_{3b}\bigg\}
\nonumber\\
{\cal A}_{85}&\sim&ist \Tr(P_{-}\fsC_{(n-1)}M_p \Gamma^{bcie}) k_{1e}\xi_{1i} k_{2c} L_2\bigg\{-2 k_{3}.\xi_{2} \xi_{3b}+2\xi_2.\xi_{3} k_{3b}\bigg\}
\nonumber\\\nonumber\\
{\cal A}_{86}&\sim& is L_1\Tr(P_{-}\fsC_{(n-1)}M_p \Gamma^{bi})\xi_{1i} \bigg\{2t k_3.\xi_2  \xi_{3b}-2t\xi_3.\xi_2 k_{3b}-2uk_1.\xi_2\xi_{3b}-4k_1.\xi_2k_2.\xi_3k_{3b}\bigg\}\nonumber\\
{\cal A}_{87}&\sim& -it L_1\Tr(P_{-}\fsC_{(n-1)}M_p \Gamma^{ai}) \xi_{1i}\bigg\{-2s k_2.\xi_3  \xi_{2a}+2s\xi_3.\xi_2 k_{2a}+2uk_1.\xi_3\xi_{2a}+4k_1.\xi_3 k_3.\xi_2 k_{2a}\bigg\}\nonumber\\
{\cal A}_{88}&\sim& i L_1 \Tr(P_{-}\fsC_{(n-1)}M_p \Gamma^{ie})\xi_{1i} k_{1e} (u\xi_2.\xi_3 +2 k_3.\xi_2 k_2.\xi_3) \bigg\{\frac{-2st+u+s+t}{(-u-\frac{1}{2})}\bigg\}
\nonumber\eeqa
where the functions
 $L_1,L_2$ are
\beqa
L_1&=&(2)^{-2(t+s+u)-1}\pi{\frac{\Gamma(-u+\frac{1}{2})
\Gamma(-s+\frac{1}{2})\Gamma(-t+\frac{1}{2})\Gamma(-t-s-u)}
{\Gamma(-u-t+1)\Gamma(-t-s+1)\Gamma(-s-u+1)}},\nonumber\\
L_2&=&(2)^{-2(t+s+u)}\pi{\frac{\Gamma(-u)
\Gamma(-s)\Gamma(-t)\Gamma(-t-s-u+\frac{1}{2})}
{\Gamma(-u-t+1)\Gamma(-t-s+1)\Gamma(-s-u+1)}}
\label{Ls}
\eeqa 
 One can simplify the S-matrix further. Indeed the 5th term  of ${\cal A}_{1}$  can be precisely cancelled by the second term of  ${\cal A}_{63}$. 
 Likewise,  the 5th term of  ${\cal A}_{3}$ is removed by  the 2nd term of ${\cal A}_{88}$. Notice that the 1st and 2nd 
 terms ${\cal A}_{42}$ are cancelled off by  the 1st and 3rd terms of 
 ${\cal A}_{87}$ accordingly. Finally the 1st and 2nd terms of  ${\cal A}_{72}$ are removed by contributions of
 the 3rd and 1st terms of ${\cal A}_{86}$ appropriately.

\vskip.08in
Let us  consider RR in terms of its field strength while 
keep track of scalar field in zero picture. The S-matrix elements of  $<V_{C^{-1}} V_{\phi^{0}} V_{A^ {-1}} V_{A^0}>$   to all orders in $\alpha'$  
has already been found in  \cite{Hatefi:2015ora} to be 
\beqa {\cal A}^{C^{-1} \phi^{0}A^ {-1}A^{0}}&=&{\cal A'}_{1}+{\cal A'}_{2}
+{\cal A'}_{3}+{\cal A'}_{4}+{\cal A'}_{5}+{\cal A'}_{6}+{\cal A'}_{7}\labell{711uui}\eeqa

where
\beqa
{\cal A'}_{1}&\!\!\!\sim\!\!\!&2^{-1/2}\xi_{1i}\xi_{2a}\xi_{3b}
\bigg[-k_{3c}k_{1d}\Tr(P_{-}\fsH_{(n)}M_p\Gamma^{bcaid})
+k_{3c}p^i\Tr(P_{-}\fsH_{(n)}M_p\Gamma^{bca})\bigg]
4(-t-s-u)L_1,
\nonumber\\
{\cal A'}_{2}&\sim&2^{-1/2}\Tr(P_{-}\fsH_{(n)}M_p \Gamma^{aid})\xi_{1i}\xi_{2a}k_{1d}
\bigg\{-2k_1.\xi_3 (ut)+2k_2.\xi_3 (st)
\bigg\}L_2\nonumber\\
{\cal A'}_{5}&\sim&2^{-1/2}\Tr(P_{-}\fsH_{(n)}M_p \gamma^{i})\xi_{1i}
\bigg\{\xi_{3}.\xi_{2}(2ts)+2k_1.\xi_3(2k_3.\xi_2)t-4u k_1.\xi_2(k_1.\xi_3)+4sk_2.\xi_3k_1.\xi_2\bigg\}L_1\nonumber\\
{\cal A'}_{4}&\sim&-2^{-1/2}(st)L_2
\bigg\{  \xi_{3b}\xi_{1i}\xi_{2a}\Tr(P_{-}\fsH_{(n)}M_p \Gamma^{bai})u+2k_3.\xi_2 k_{1d}\xi_{1i}\xi_{3b}\Tr(P_{-}\fsH_{(n)}M_p \Gamma^{bid})\nonumber\\&&
-\Tr(P_{-}\fsH_{(n)}M_p \Gamma^{cid})k_{1d}k_{3c}\xi_{1i}(2\xi_2.\xi_3)
\bigg\}
\nonumber\\
{\cal A'}_{3}&\sim&-2^{-1/2}\xi_{1i}k_{3c}
\bigg\{-2k_1.\xi_2 \xi_{3b}\Tr(P_{-}\fsH_{(n)}M_p \Gamma^{bci})(us)+2k_1.\xi_3 \xi_{2a}\Tr(P_{-}\fsH_{(n)}M_p \Gamma^{cai})(ut)
\bigg\}L_2\nonumber\\
{\cal A'}_{6}&\sim&2^{-1/2}(st) p^i \xi_{1i}
\bigg\{ 2k_3.\xi_2 \Tr(P_{-}\fsH_{(n)}M_p \gamma^{b})\xi_{3b}-2\xi_3.\xi_2 \Tr(P_{-}\fsH_{(n)}M_p \gamma^{c})k_{3c}
\bigg\}L_2\nonumber\\
{\cal A'}_{7}&\sim&2^{-1/2} p^i \xi_{1i}\Tr(P_{-}\fsH_{(n)}M_p \gamma^{a})\xi_{2a}
\bigg\{ 2k_1.\xi_3 (ut)-2\xi_3.k_2(st)\bigg\}L_2\labell{483}\eeqa

Finally if we  deal with RR in terms of its field strength and consider the picture of scalar field in (-1) picture, 
then the S-matrix elements $<V_{C^{-1}} V_{\phi^{-1}} V_{A^ {0}} V_{A^0}>$  to all orders in $\alpha'$ can be explored
\cite{Hatefi:2012ve} as follows
\beqa {\cal A}^{C^{-1} \phi^{-1}A^ {0}A^0}&=&{\cal A''}_{1}+{\cal A''}_{2}+{\cal A''}_{3}+{\cal A''}_{4}
+{\cal A''}_{5}\labell{712u}\eeqa
where
\beqa
{\cal A''}_{1}&\!\!\!\sim\!\!\!&-2^{-1/2}\xi_{1i}\xi_{2a}\xi_{3b}
\bigg[k_{3d}k_{2c}\Tr(P_{-}\fsH_{(n)}M_p\Gamma^{bdaci})
\bigg]
4(-t-s-u)L_1,
\nonumber\\
{\cal A''}_{2}&\sim&2^{-1/2}\Tr(P_{-}\fsH_{(n)}M_p \Gamma^{bdi})\xi_{1i}\xi_{3b}k_{3d}
\bigg\{2k_1.\xi_2 (us)-2k_3.\xi_2 (st)
\bigg\}L_2\\
{\cal A''}_{3}&\sim&-2^{-1/2}\Tr(P_{-}\fsH_{(n)}M_p \Gamma^{aci})\xi_{1i}\xi_{2a}k_{2c}
\bigg\{-2k_2.\xi_3 (st)+2k_1.\xi_3 (ut)
\bigg\}L_2\nonumber\\
{\cal A''}_{4}&\sim&-2^{-1/2}(st)L_2
\bigg\{ \xi_{3b}\xi_{1i}\xi_{2a}\Tr(P_{-}\fsH_{(n)}M_p \Gamma^{bai})u +2k_2.\xi_3 k_{3d}\xi_{1i}\xi_{2a}\Tr(P_{-}\fsH_{(n)}M_p \Gamma^{dai})
\nonumber\\&&+2k_3.\xi_2 k_{2c}\xi_{1i}\xi_{3b}\Tr(P_{-}\fsH_{(n)}M_p \Gamma^{bci})-\Tr(P_{-}\fsH_{(n)}M_p \Gamma^{dci})k_{3d}k_{2c}\xi_{1i}(2\xi_2.\xi_3)
\bigg\}\nonumber\\
{\cal A''}_{5}&\sim&2^{-1/2}\Tr(P_{-}\fsH_{(n)}M_p \gamma^{i})\xi_{1i}
\bigg\{
\xi_{3}.\xi_{2}(2ts)+2k_1.\xi_3(2k_3.\xi_2)t-4u k_1.\xi_2(k_1.\xi_3)\nonumber\\&&
+4sk_2.\xi_3k_1.\xi_2\bigg\}L_1\nonumber
\labell{48333}\eeqa
All  singularity structure and contact interaction comparisons for both symmetric pictures have 
already been done in \cite{Hatefi:2015ora}.
To be able to explore all the bulk singularity structures, in the following sections we first try to compare 
all order $\alpha'$ singularity structures in symmetric and asymmetric picture and afterwards we
carry out the comparisons at the level of contact terms to see whether or not some new SYM couplings can be discovered.

\section{All order  singularity comparisons between
\\ 
$<V_{C^{-2}}V_{\phi^{0}} V_{A^{0}} V_{A^{0}}>$ and  $<V_{C^{-1}}V_{\phi^{0}} V_{A^{-1}} V_{A^{0}}>$}

In this section we first work with $<V_{C^{-2}}V_{\phi^{0}} V_{A^{0}} V_{A^{0}}>$ S-matrix, trying to reconstruct all the singularity structures of $<V_{C^{-1}}V_{\phi^{0}} V_{A^{-1}} V_{A^{0}}>$ and then start exploring new bulk singularity structures. If we add the 1st terms of ${\cal A}_{3}$ and ${\cal A}_{88}$ of asymmetric picture \reef{48ii},
then we obtain the following terms 
\beqa
-2ist  k_{1e} \Tr(P_{-}\fsC_{(n-1)}M_p \Gamma^{ei})\xi_{1i}\xi_2.\xi_{3}   L_1 \label{yy13}\eeqa
Now one needs to actually consider the sum of  \reef{yy13} with the 2nd terms of ${\cal A}_{86},{\cal A}_{87}$ accordingly to gain the following term 
\beqa
-2ist  (k_1+k_2+k_3)_{c}\Tr(P_{-}\fsC_{(n-1)}M_p \Gamma^{ci})\xi_{1i}\xi_2.\xi_{3}   L_1 \label{yy11}\eeqa
If we apply the momentum conservation to \reef{yy11}  and use $p\fsC=\fsH$ then we  immediately reveal that,  this term is 
exactly the first term ${\cal A'}_{5}$ of symmetric picture \reef{711uui}. Likewise, we need to add the 2nd terms of  ${\cal A}_{3}, {\cal A}_{43}$  and the 3rd term ${\cal A}_{72}$ to be able to  derive  the following term
\beqa
-4iu  (k_1+k_2+k_3)_{c}\Tr(P_{-}\fsC_{(n-1)}M_p \Gamma^{ic})\xi_{1i} k_1.\xi_2 k_1.\xi_{3}   L_1 \nonumber\eeqa
which is exactly the the 3rd term ${\cal A'}_{5}$. If we also add the last terms of  ${\cal A}_{72}, {\cal A}_{87}$
 and the 3rd term  ${\cal A}_{3}$, we get to 
\beqa
4it  (k_1+k_2+k_3)_{c}\Tr(P_{-}\fsC_{(n-1)}M_p \Gamma^{ic})\xi_{1i} k_1.\xi_3 k_3.\xi_{2}   L_1 \nonumber\eeqa
which is the 2nd  term  ${\cal A'}_{5}$. Eventually, by adding the fourth terms of 
 ${\cal A}_{3}, {\cal A}_{86}$  and the 1st term  ${\cal A}_{43}$, we explore the last term 
 ${\cal A'}_{5}$ as follows
\beqa
4is (k_1+k_2+k_3)_{c}\Tr(P_{-}\fsC_{(n-1)}M_p \Gamma^{ic})\xi_{1i} k_1.\xi_2 k_2.\xi_{3}   L_1 \nonumber\eeqa
Note that the 1st term  ${\cal A}_{5}$ produces exactly the first term ${\cal A'}_{3}$. By applying momentum conservation to the last term 
 ${\cal A}_{62}$ , apart from reconstructing the last term  ${\cal A'}_{4}$, we  generate the 2nd term  ${\cal A'}_{6}$ as well.
 
 \vskip.1in

Applying $k_{2c}=-(k_1+k_3+p)_c$  to both terms of ${\cal A}_{41}$  and to the  2nd term ${\cal A}_{83}$ and keeping in mind the
antisymmetric property of $\epsilon$ tensor,  
not only both terms of ${\cal A'}_{2}$ but also the 2nd term ${\cal A'}_{3}$ 
are precisely reconstructed, while the extra terms coming from  momentum conservation can be precisely
cancelled off, where the 2nd term  ${\cal A}_{84}$ is also taken into account. Having applied  momentum conservation to  the 3rd term  ${\cal A}_{62}$, one would find the following term
  \beqa
 -2k_3.\xi_2 ip.\xi_1\Tr(P_{-}\fsC_{(n-1)}M_p \Gamma^{ba})  (ts L_2 )   \xi_{3b}  (k_1+p+k_3)_{a}
 \label{esi98001}\eeqa
 where the 1st term in \reef{esi98001} produces the 2nd term  ${\cal A'}_{4}$, its second term generates the 1st term  of ${\cal A'}_{6}$  and essentially its 3rd term will be cancelled by the 2nd term ${\cal A}_{5}$ of asymmetric amplitude.
 
 \section{ Bulk singularity structures of 
 $<V_{C^{-2}}V_{\phi^{0}} V_{A^{0}} V_{A^{0}}>$ }

We start to address all order bulk singularities of asymmetric amplitude. Indeed  we consider the other 
terms in the S-matrix of \reef{48ii} that have not been taken into account in $<V_{C^{-1}}V_{\phi^{0}} V_{A^{-1}} V_{A^{0}}>$ . 
To deal with those terms,  we need to add the first terms  of ${\cal A}_{1}, {\cal A}_{63}$   
to get to the following terms
 \beqa
 -2st ip.\xi_1 \xi_2.\xi_3  \Tr(P_{-}\fsC_{(n-1)}M_p )L_1  \label{yy22}\eeqa
 One also needs to relate \reef{yy22} to the other terms that have been leftover in ${\cal A}_{1}$ of  \reef{48ii} (all its terms except its 5th term), to eventually explore  new bulk singularity structures in asymmetric picture as follows
 
 \beqa
 i p.\xi_1 \Tr(P_{-}\fsC_{(n-1)}M_p) L_1
\bigg[-2st\xi_{3}.\xi_{2} +4u k_1.\xi_{3} k_1.\xi_2  -4t k_3.\xi_{2} k_1.\xi_3 -4s  k_2.\xi_{3} k_1.\xi_2 \bigg].
\label{ea11}\eeqa
Let us produce bulk poles in an EFT.
 
\subsection {An infinite number of $(s+t+u)$-channel bulk singularities }
Let us construct all these new $(s+t+u)$- channel bulk singularity structures in an effective field theory. The expansion of  $L_1$ and its all coefficients is given in \cite{Hatefi:2015ora}.
We extract the trace and just focus on the part of the expansion that involves all  poles  as below
\beqa
&&8i\pi^3\mu_p p.\xi_{1} \eps^{a_{0}\cdots a_{p}} 
C_{a_0\cdots a_{p}}\frac{1}{(p+1)!(s+t+u)}
\sum_{n,m=0}^{\infty}c_{n,m}(s^{m}t^{n}+s^{n}t^{m})\nonumber\\&&
\bigg[-2st\xi_{2}.\xi_3-4t k_1.\xi_3 k_3.\xi_2-4s k_1.\xi_2 k_2.\xi_3+4u k_1.\xi_2 k_1.\xi_3\bigg]\Tr(\lam_1\lam_2\lam_3)
\label{amphigh87369}\eeqa

If we take into account the following effective field theory amplitude
\beqa
V_{\alpha}^{i}(C_{p+1},\phi)G_{\alpha\beta}^{ij}(\phi)V_{\beta}^{j}(\phi,\phi_1,
A_2,A_3)\label{vienna1}\eeqa
and scalar field 's propagator  ($\frac{(2\pi\alpha')^2} {2}D^a\phi^iD_a\phi_i$) as well as the following coupling 

\beqa
(2\pi\alpha')i\mu_p\int d^{p+1}\sigma {1\over (p+1)!}
(\veps)^{a_0\cdots a_{p}} D_{a_{0}}\phi_i C^{i}_{a_1\cdots a_{p}} \labell{vie2}\eeqa
then one would be able to find out the vertex of 
 of $V_{\alpha}^{i}(C_{p+1},\phi)$ as follows

\beqa
G_{\alpha\beta}^{ij}(\phi) &=&\frac{-i\delta_{\alpha\beta}\delta^{ij}}{T_p(2\pi\alpha')^2
k^2}=\frac{-i\delta_{\alpha\beta}\delta^{ij}}{T_p(2\pi\alpha')^2
(t+s+u)},\nonumber\\
V_{\alpha}^{i}(C_{p+1},\phi)&=&i(2\pi\alpha')\mu_p\frac{1}{(p+1)!}(\veps)^{a_0\cdots a_{p}}
 k_{a_{0}} C^{i}_{a_1\cdots a_{p}}\Tr(\lambda_{\alpha}).
\labell{Fey}
\eeqa
Now we employ all order $\alpha'$ two gauge field-two scalar field 
SYM couplings  that have been derived in  \cite{Hatefi:2012ve} as follows
\beqa
(2\pi\alpha')^4\frac{1}{ 2 \pi^2}T_p\left(\alpha'\right)^{n+m}\sum_{m,n=0}^{\infty}(\cL_{1}^{nm}+\cL_{2}^{nm}+\cL_{3}^{nm}),\labell{highder}\eeqa
\beqa
&&\cL_{1}^{nm}=-
\Tr\left(\frac{}{}a_{n,m}\cD_{nm}[D_a \phi^i D^b \phi_i F^{ac}F_{bc}]+ b_{n,m}\cD'_{nm}[D_a \phi^i F^{ac} D^b \phi_i F_{bc}]+h.c.\frac{}{}\right),\nonumber\\
&&\cL_{2}^{nm}=-\Tr\left(\frac{}{}a_{n,m}\cD_{nm}[D_a \phi^i D^b \phi_i F_{bc}F^{ac}]+\frac{}{}b_{n,m}\cD'_{nm}[D_a \phi^i F_{bc} D^b \phi_i F^{ac}]+h.c.\frac{}{}\right),\nonumber\\
&&\cL_{3}^{nm}=\frac{1}{2}\Tr\left(\frac{}{}a_{n,m}\cD_{nm}[D_a \phi^i D^a \phi_i F^{bc}F_{bc}]+\frac{}{}b_{n,m}\cD'_{nm}[D_a \phi^i F_{bc} D^a \phi_i F^{bc}]+h.c\frac{}{}\right),\nonumber\eeqa
for which special definitions of $\cD_{nm}(EFGH) ,\cD'_{nm}(ABCD)$ for all higher derivative operators 
are appeared in \cite{Hatefi:2010ik}.

\vskip.1in

To be able to actually produce all $(t+s+u)$-channel bulk singularities, one needs to explore $ V_{\beta}^{j}(\phi,\phi_1,
A_2,A_3)$  from  \reef{highder}. Some remarks are also worth pointing out.  The momentum conservation 
$(k_1+k_2+_3)^{a}=-p^a$ and $p_{a_{0}}  C^{i}_{a_1\cdots a_{p}}=\mp p_{i}  C_{a_0\cdots a_{p}}$ have been 
taken. $k$ is the momentum of off-shell Abelian scalar field. The orderings of $\Tr(\lam_1\lambda_{\beta}\lam_2\lam_3),
\Tr(\lambda_{\beta}\lam_1\lam_2\lam_3)$ are also considered in such a way that one is able to find 
$ V_{\beta}^{j}(\phi,\phi_1,A_2,A_3)$  as follows


\beqa
V_{\beta}^{j}(\phi,\phi_1,
A_2,A_3)&=&\xi_{1}^j\frac{(2\pi\alpha')^4T_{p}}{2\pi^2}\bigg(\frac{st}{2}\xi_{2}.\xi_3+t k_1.\xi_3 k_3.\xi_2+s k_1.\xi_2 k_2.\xi_3-u k_1.\xi_2 k_1.\xi_3\bigg)(\alpha')^{n+m}\nonumber\\&&\times
\bigg((k_3\inn k_1)^m(k_1\inn k_2)^n+(k_3\inn k)^m(k_2\inn k)^n
+(k_1\inn k_3)^n(k_1\inn k_2)^m\nonumber\\&&+(k\inn k_3)^n (k\inn k_2)^m\bigg)(a_{n,m}+b_{n,m})\Tr(\lam_1\lam_2\lam_3\lambda_{\beta})
\labell{verppaa}\eeqa
Note that all   $a_{n,m},b_{n,m}$  coefficients are explored in \cite{Hatefi:2010ik}  such as
  \beqa
a_{0,0}=-\frac{\pi^2}{6},\,b_{0,0}=-\frac{\pi^2}{12},a_{1,0}=2\z(3),\,a_{0,1}=0,\,b_{0,1}=-\z(3),a_{1,1}=a_{0,2}=-7\pi^4/90,\nonumber\\
a_{2,2}=(-83\pi^6-7560\z(3)^2)/945,b_{2,2}=-(23\pi^6-15120\z(3)^2)/1890\nonumber\\
\,a_{2,0}=-4\pi^4/90,\,b_{1,1}=-\pi^4/180,\,b_{0,2}=-\pi^4/45\nonumber\eeqa
with $b_{n,m}$ 's become symmetric.
Substituting  \reef{verppaa} and \reef{Fey} into effective field theory amplitude \reef{vienna1}, we obtain all order $\alpha'$ bulk singularity structures in the field theory side as
\beqa
&&16\pi\mu_p\frac{\eps^{a_{0}\cdots a_{p}} p.\xi_{1}
C_{a_0\cdots a_{p}}}{(p+1)!(s+t+u)}\Tr(\lam_1\lam_2\lam_3)
\sum_{n,m=0}^{\infty} (a_{n,m}+b_{n,m})[s^{m}t^{n}+s^{n}t^{m}]\nonumber\\&&
\bigg[2st\xi_{2}.\xi_3+4t k_1.\xi_3 k_3.\xi_2+4s k_1.\xi_2 k_2.\xi_3-4u k_1.\xi_2 k_1.\xi_3\bigg]
\label{amphigh8}\eeqa
In order to  show that we have exactly explored all order  $(t+s+u)$-channel bulk singularities of asymmetric string amplitude, we just 
remove the common factors and start comparing  at each order of $\alpha'$ both string and field theory amplitudes \reef{amphigh87369}
and \reef{amphigh8} accordingly. 

\vskip.1in

At zeroth order for $n=m=0$, the EFT amplitude  \reef{amphigh8} produces   the following coefficient
$-4(a_{0,0}+b_{0,0})=\pi^2$ and
the  string amplitude \reef{amphigh87369}
regenerates $(2\pi^2 c_{0,0})=\pi^2$. At 1st order  \reef{amphigh8} will be led to  
$-(a_{1,0}+a_{0,1}+b_{1,0}+b_{0,1})(s+t)=0$, and
 string amplitude carries $\pi^2(c_{1,0}+c_{0,1})(s+t)=0$ coefficient. At  $(\alpha')^2$ order, \reef{amphigh8} predicts 
 the coefficient $-4(a_{1,1}+b_{1,1})st-2(a_{0,2}+a_{2,0}+b_{0,2}+b_{2,0})[s^2+t^2]=\frac{\pi^4}{3}(st)+\frac{\pi^4}{3}(s^2+t^2)$
and string amplitude is also equivalent to  $\pi^2 [c_{1,1}(2st)+(c_{2,0}+c_{0,2})(s^2+t^2)]$,  which is precisely the same as EFT coefficient. One can get the same results at all orders of $\alpha'$ (see \cite{Hatefi:2012ve} for $(\alpha')^3,(\alpha')^4$ comparisons).
\vskip.1in

Hence in this section by introducing a new coupling \reef{vie2} we were precisely able to obtain all order  $(t+s+u)$-channel 
bulk singularity structures  in both string and field theory sides. Let us concentrate on the other bulk singularity structures.

 \subsection {An infinite number of $t,s$-channel bulk singularity structures }

Let us proceed with the 2nd terms  of ${\cal A}_{71}, {\cal A}_{82}$ in asymmetric picture. If we apply momentum conservation  $(k_3+p)_{a}=-(k_1+k_2)_{a}$ to those terms, we then obtain 
  \beqa
  2i k_1.\xi_2 suL_2  \xi_{3b}\xi_{1i} k_{3d}\Tr(P_{-}\fsC_{(n-1)}M_p \Gamma^{bdai})  (p+k_3)_{a} \label{121908}\eeqa 
  the second term in \reef{121908} has evidently zero contribution to S-matrix, given the fact that  \reef{121908}
  is symmetric under exchanging $k_{3d}, k_{3a}$ while simultaneously is antisymmetric under $\epsilon$ tensor so 
  the result for the 2nd term of \reef{121908} is zero.  Its first term includes an infinite number 
  of t-channel  singularities. Indeed these t-channel singularities are the same poles that have been 
  shown up in the  1st terms   ${\cal A''}_{2}$ of \reef{712u}  and ${\cal A'}_{3}$ of \reef{711uui} 
  and these t-channel poles have already been produced in \cite{Hatefi:2012ve} as well.

\vskip.2in

 However, the point of this section is that, besides those singularities,  this S-matrix  includes an infinite t-channel bulk 
 singularities where we just demonstrate them right now. It is also important to highlight  that  the 1st term 
 ${\cal A}_{5}$ of asymmetric amplitude is a new bulk singularity structure  as follows
 \beqa
  -2ius k_1.\xi_{2}  k_{3d}\xi_{3b}p.\xi_1\Tr(P_{-}\fsC_{(n-1)}M_p \Gamma^{bd}) L_2 \label{jji156}\eeqa 
  that has an infinite number of t-channel bulk singularity structures where we reconstruct them in the following.
Note that due to symmetries (as can be seen later on), the amplitude is antisymmetric under exchanging either
$2 \leftrightarrow 3 $ or $t \leftrightarrow s$, therefore we just regenerate all infinite t-channel bulk 
singularities and finally by changing all the momenta and polarizations of two gauge fields one can produce all
s-channel bulk poles as well.
\vskip.1in

Inserting just singularity  contributions of  $(us L_2)$ expansion to \reef{jji156}, 
one finds out all infinite t-channel bulk singularities of asymmetric string amplitude (where $(\pi)^{1/2} \mu_p$ is a normalization 
constant  of the amplitude) as follows
\beqa
-ip.\xi_1 \frac{(2\pi\alpha')^2}{ (p-1)!}\mu_p  \xi_{3b}k_{3d}\eps^{a_{0}\cdots a_{p-2}bd}C_{a_{0}\cdots a_{p-2}}
\sum_{n=-1}^{\infty}\frac{1}{t}{b_n(u+s)^{n+1}}(2k_1.\xi_2)\label{UI99}
\eeqa
We also take into account the following  sub-amplitude in an effective field theory 
\beqa
{\cal A}&=&V^i_{\alpha}(C_{p-1},A_3,\phi)G^{ij}_{\alpha\beta}(\phi)V^j_{\beta}(\phi,A_2,\phi_1)\label{vvcx33}
\eeqa 
 The kinetic term of scalars $ \frac{(2\pi\alpha')^2 }{2}D^a\phi^i D_a\phi_i$  has been taken to derive  the following vertex and propagator
 \beqa
V^j_{\beta}(\phi,A_2,\phi_1)&=&-2ik_1.\xi_2 (2\pi\alpha')^2 T_p  \xi_1^j\Tr(\lam_1\lam_2\lam_\beta)
\labell{fvertex78}\\
({G}^{\phi})^{ij}_{\alpha\beta}&=&-\frac{i\delta^{ij}\delta^{\alpha\beta}}{(2\pi\alpha')^2T_p t}
\nonumber
\eeqa
One also needs to apply Taylor expansion of a real scalar field through the Chern-Simons action
$\bigg({i}(2\pi\alpha')^2\mu_p\int d^{p+1}\s {1\over (p-1)!}   \prt_i C^{(p-1)}\wedge F\phi^i \bigg)$ 
to obtain $V^i_{\alpha}(C_{p-1},A_3,\phi)$ as

\beqa
V^i_{\alpha}(C_{p-1},A_3,\phi)&=&\frac{i (2\pi\alpha')^2\mu_p}{(p-1)!}(\epsilon)^{a_0\cdots a_p}C_{a_0\cdots a_{p-2}} p^i \xi_{3a_{p}}k_{3a_{p-1}}\Tr(\lam_3\lambda_\alpha)\label{vvcx22}\eeqa
where $V^j_{\beta}(\phi,A_3,\phi_1)$ was obtained from  scalar fields 's kinetic term in DBI effective action, and it receives no correction as all the kinetic terms have already been fixed in the action. Therefore to produce all infinite t-channel bulk singularity structures, we just impose all order corrections to the following mixed coupling as below 
\beqa
 \sum_{n=-1}^{\infty} b_n (\alpha')^{(n+1)}\mu_p\int d^{p+1}\s {1\over (p-1)!}
\partial_{i}C_{p-1}\wedge D_{a_{1}}...D_{a_{n+1}} F D^{a_{1}}...D^{a_{n+1}}\phi^i
\label{vvcx}
\eeqa

Having regarded \reef{vvcx}, one could reveal  all order extended vertex operator as follows 
\beqa
V^i_{\alpha}(C_{p-1},A_3,\phi)&=&\frac{i (2\pi\alpha')^2\mu_p}{(p-1)!}(\epsilon)^{a_0\cdots a_p} 
C_{a_0\cdots a_{p-2}} p^i \xi_{3a_{p}}k_{3a_{p-1}}\nonumber\\&&\times\Tr(\lam_3\lambda_\alpha)\sum_{n=-1}^{\infty}
b_n(\alpha'k_3.k)^{n+1}
\label{imb1789}\eeqa
where $\sum_{n=-1}^{\infty}b_n(\alpha'k_3.k)^{n+1}=\sum_{n=-1}^{\infty}b_n(u+s)^{n+1}$ is employed. Replacing \reef{imb1789} and \reef{fvertex78} to  \reef{vvcx33}, we are able to explore all order t-channel bulk singularities
of string amplitude  as follows
\beqa
ip.\xi_1 \frac{(2\pi\alpha')^2}{ (p-1)!}\mu_p  \xi_{3a_{p-1}}k_{3a_{p}}\eps^{a_{0}\cdots a_{p}}
C_{a_{0}\cdots a_{p-2}}\sum_{n=-1}^{\infty}\frac{1}{t}{b_n(u+s)^{n+1}}(2k_1.\xi_2)\label{UI9944}
\eeqa
which are precisely the bulk singularities that have been found in \reef{UI99}, so we were able to
reconstruct them in an effective theory as well.
Finally let us deal with all s-channel bulk singularities. We need to actually consider the sum of  the first  term
${\cal A}_{41}$
with the 2nd term  ${\cal A}_{83}$  and make use of momentum conservation along the world volume 
of brane $(k_1+k_3)_{e}=-(k_2+p)_e$ to derive the following term 
\beqa
-2iut  k_{1}.\xi_3 \Tr(P_{-}\fsC_{(n-1)}M_p \Gamma^{acei})\xi_{1i} \xi_{2a}  k_{2c}  (k_2+p)_e L_2 \label{mmnn}\eeqa
the first term in \reef{mmnn} has no contribution to S-matrix, because the whole \reef{mmnn} 
is symmetric under exchanging $k_{2c}, k_{2e}$ and also is antisymmetric under $\epsilon$ tensor so 
the result for the first term of \reef{mmnn} is zero. Its second term $(p\fsC=\fsH)$ involves an infinite 
number of s-channel singularities  which are precisely the 2nd term of
${\cal A''}_{3}$ and these s-channel poles have already been produced in \cite{Hatefi:2012ve}.\footnote{On the other hand, applying the following Bianchi identity 
\beqa
\xi_{1i} \xi_{2a} \bigg( \epsilon^{a_{0}...a_{p-2}da} p_{d} H^i_{a_0...a_{p-2}}+p^i \epsilon^{a_0...a_{p-1}a} H_{a_{0}...a_{p-1}}\bigg)=0\nonumber\eeqa
and considering the sum of the 1st term of ${\cal A'}_{2}$, the 2nd term of ${\cal A'}_{3}$ and 1st term  of ${\cal A'}_{7}$ of \reef{711uui} will give rise the same s-channel singularities as follows

\beqa
 2iut  k_{1}.\xi_3 \Tr(P_{-}\fsH_{(n-1)}M_p \Gamma^{aid})\xi_{1i} \xi_{2a}    (k_2+p)_{d} L_2 \nonumber\\&&
\label{mmnn99}\eeqa}
\vskip.1in

 It is also worth emphasizing the important point  as follows. The 1st term   ${\cal A}_{2}$ of 
 asymmetric amplitude  \reef{48ii} is also a new bulk singularity structure as follows
 \beqa
  2iut k_1.\xi_{3}  k_{2c}\xi_{2a}p.\xi_1\Tr(P_{-}\fsC_{(n-1)}M_p \Gamma^{ac}) L_2 \label{jji1}\eeqa 
  that can be re-derived in an EFT as shown for an infinite number of t-channel bulk poles of  this section.

 \subsection {An infinite number of $u$-channel singularities }

In this section, we add the 2nd terms  ${\cal A}_{62}, {\cal A}_{2}$ of  \reef{48ii} and apply momentum
conservation along the brane to get to the following terms
 \beqa
  -its p.\xi_1 (2k_2.\xi_3) \Tr(P_{-}\fsC_{(n-1)}M_p \Gamma^{ac}) L_2  \xi_{2a} (-p-k_1)_{c} \label{op09}\eeqa
  where the first term in \reef{op09} reconstructs the 2nd term ${\cal A'}_{7}$    
  and its second term remains to be  explored as an extra u-channel pole. Adding the 1st terms  ${\cal A}_{71}, {\cal A}_{85}$ we derive the following u-channel bulk singularities
  \beqa
  2i k_3.\xi_2 tsL_2  \xi_{3b}\xi_{1i} k_{1e}\Tr(P_{-}\fsC_{(n-1)}M_p \Gamma^{bdie})  (p+k_1)_{d} \label{121800}\eeqa 
 Clearly the second term  in \reef{121800}  has  no contribution to the S-matrix at all, because the whole \reef{121800} 
 is symmetric under exchanging $k_{1d}, k_{1e}$ and simultaneously is antisymmetric under $\epsilon$ tensor, 
 so the result is zero. The first term in \reef{121800} includes an infinite number of new u-channel bulk 
 singularities that would be generated in an effective field theory in the following section.
Note that the 2nd term of ${\cal A}_{85}$ needs to be taken as follows
 \beqa
2i\xi_2.\xi_{3} stL_2   \Tr(P_{-}\fsC_{(n-1)}M_p \Gamma^{bcie}) k_{1e}\xi_{1i} k_{2c}  k_{3b} \label{maz}\eeqa
which has contribution to  an infinite number of u-channel gauge poles as well. 
We now collect all the above terms, extracting the traces and write down all u-channel poles as 
\beqa
 &&i\mu_p\xi_{1i}k_{1e}\bigg[\eps^{a_{0}\cdots a_{p-3}ebd} p_d
\bigg(2k_3.\xi_2\xi_{3b} C^i{}_{a_0\cdots a_{p-3}} -2\xi_3.\xi_2 k_{3b} C^i{}_{a_0\cdots a_{p-3}}\bigg)
\nonumber\\&&-2k_2.\xi_3\xi_{2a} p^i C_{a_0\cdots a_{p-2}}\eps^{a_{0}\cdots a_{p-2}ea}\bigg]
(2\pi\alpha')^{2}\frac{1}{(p-1)!u}\times\sum_{n=-1}^{\infty}b_n (s+t)^{n+1}
\labell{amp00128}\eeqa
where the following field theory amplitude must be considered

\beqa
{\cal A}&=&V^a_{\alpha}(C_{p-1},\phi_1,A)G^{ab}_{\alpha\beta}(A)V^b_{\beta}(A,A_2,A_3),\label{amp0098}
\eeqa
where gauge field propagator and the other vertex operator are read off
\beqa
V^b_{\beta}(A,A_2,A_3)&=&-iT_p(2\pi\alpha')^{2}\Tr(\lam_2\lam_3\lambda_\beta)\bigg[2k_2.\xi_3\xi_2^b
-2 k_3.\xi_2\xi_3^b+\xi_3.\xi_2(k_3-k_2)^b\bigg],\nonumber\\
G_{\alpha\beta}^{ab}(A)&=&\frac{i\delta_{\alpha\beta}\delta^{ab}}{(2\pi\alpha')^2 T_p
u},\label{amp0017}
\eeqa
One also needs  to deal with Taylor expansion of a real scalar field to Chern-Simons coupling as we mentioned it
earlier on, so that by considering its all order corrections as  \reef{vvcx}, one is able to re-derive  all order vertex operator of $V^a_{\alpha}(C_{p-1},\phi_1,A)$ as follows

\beqa
V^a_{\alpha}(C_{p-1},\phi_1,A)&=&\frac{i(2\pi\alpha')^2\mu_p}{(p-1)!}(\veps)^{a_0\cdots a_{p-1}a}  C_{a_0\cdots a_{p-2}} p.\xi_{1}k_{a_{p-1}}\Tr(\lam_1\lambda_\alpha)\sum_{n=-1}^{\infty}b_n(t+s)^{n+1}.\nonumber\eeqa
where $k$ is now the momentum of off-shell gauge field $k^a=-(k_2+k_3)^a=(p+k_1)^a$.
Replacing  the above vertex and \reef{amp0017} into  \reef{amp0098}, we would be able to exactly reconstruct all order u-channel gauge field poles of string amplitude \reef{amp00128} as well.
\vskip.1in

 Finally let us  make comparisons to all order $\alpha'$ contact interactions of both symmetric and asymmetric pictures
 of this S-matrix.
 
 \section{Contact terms}
 
First of all, the 1st term  ${\cal A}_{62}$ of \reef{48ii} precisely produces  the 1st term ${\cal A'}_{4}$ of  \reef{711uui}.
Consider ${\cal A}_{81}$  and apply momentum conservation along the world volume of brane 
$k_{2c}=-(p+k_1+k_3)_c$ to this term, by doing so, making use of
the property of antisymmetric $\epsilon $ tensor and $pC=H$, one is able to produce the first contact term 
${\cal A'}_{1}$ of \reef{711uui}. Finally to regenerate the 2nd  term  of ${\cal A'}_{1}$, one needs to apply once more the momentum conservation 
$k_{2c}=-(p+k_1+k_3)_c$ to ${\cal A}_{61}$ and uses the antisymmetric property
of $\epsilon $ tensor where this leads to a new sort of bulk contact interaction as follows 
\beqa
4i\Tr(P_{-}\fsC_{(n-1)}M_p \Gamma^{dbac})\xi_{2a}\xi_{3b} p.\xi_1 k_{3d} k_{1c} L_1 (-t-s-u),\label{fwf1}\eeqa
Ultimately, considering the sum of the 1st terms  of ${\cal A}_{82}, {\cal A}_{83}, {\cal A}_{84}$ of 
asymmetric picture, takes us to another sort of new contact interaction which has the following structure 
\beqa
 i  p_d \Tr(P_{-}\fsC_{(n-1)}M_p \Gamma^{bdai})\xi_{1i} \xi_{2a} \xi_{3b}   ustL_2 \label{fwf22}\eeqa
Note that the above term is antisymmetric under interchanging the gauge fields. 
The method of finding all order $\alpha'$ higher derivative corrections to BPS string amplitudes 
has been  explained in sections 4,5 of \cite{Hatefi:2015ora}. To shorten the paper we just refer the interested
reader to those concrete explanations in \cite{Hatefi:2015ora}.
Indeed one could easily follow  \cite{Hatefi:2015ora} and start to generate these new contact interactions to all orders in an EFT as well.
This ends our goals of getting new bulk singularity structures as well as new couplings in string theory.
\vskip.1in

Lastly, it would be interesting to see what happens to the other asymmetric higher point  string amplitudes
of the mixed C-field and an even number of transverse scalar fields and at least a gauge field.
We hope to successfully overcome this problem and various other issues including the beautiful mathematical structures
behind string theory amplitudes in near future.

\renewcommand{\theequation}{A.\arabic{equation}}
 \setcounter{equation}{0}
  \section*{Appendix : Notations
  }

 In this appendix we would like to mention the notations  used for the momenta, positions and indices.
 The lowercase
Greek indices take values for 10 dimensional spacetime, as 
\beqa
\mu,\nu = 0, 1,..., 9
\eeqa
 The world volume indices run as $a, b, c = 0, 1,..., p$  and transverse directions of the brane are represented by $i,j = p + 1,...,9$. The Doubling trick can be taken as  follows  
\begin{displaymath}
\tilde{X}^{\mu}(\bar{z}) \rightarrow D^{\mu}_{\nu}X^{\nu}(\bar{z}) \ ,
\spa
\tilde{\psi}^{\mu}(\bar{z}) \rightarrow
D^{\mu}_{\nu}\psi^{\nu}(\bar{z}) \ ,
\spa
\tilde{\phi}(\bar{z}) \rightarrow \phi(\bar{z})\,, \mand
\tilde{S}_{\al}(\bar{z}) \rightarrow M_{\al}{}^{\be}{S}_{\be}(\bar{z})
 ,
\end{displaymath}
where
\begin{displaymath}
D = \left( \begin{array}{cc}
-1_{9-p} & 0 \\
0 & 1_{p+1}
\end{array}
\right) \ ,\,\, \mand
M_p = \left\{\begin{array}{cc}\frac{\pm i}{(p+1)!}\ga^{a_{1}}\ga^{a_{2}}\ldots \ga^{a_{p+1}}
\eps_{a_{1}\ldots a_{p+1}}\,\,\,\,{\rm for\, p \,even}\\ \frac{\pm 1}{(p+1)!}\ga^{a_{1}}\ga^{a_{2}}\ldots \ga^{a_{p+1}}\ga_{11}
\eps_{a_{1}\ldots a_{p+1}} \,\,\,\,{\rm for\, p \,odd}\end{array}\right.
\end{displaymath}
 Propagators  for the entire world-sheet fields $X^{\mu},\psi^{\mu}, \phi$ are
\begin{eqnarray}
\lan X^{\mu}(z)X^{\nu}(w)\ran & = & -\frac{\alpha'}{2}\eta^{\mu\nu}\log(z-w) , \non \\
\lan \psi^{\mu}(z)\psi^{\nu}(w) \ran & = & -\frac{\alpha'}{2}\eta^{\mu\nu}(z-w)^{-1} \ ,\non \\
\lan\phi(z)\phi(w)\ran & = & -\frac{\alpha'}{2}\log(z-w) \ .
\labell{prop}\end{eqnarray}
We also introduce
\beqa
 x_{4}\equiv\ z=x+iy\quad,\quad x_{5}\equiv\bz=x-iy
\eeqa
and SL(2,R) symmetry has been fixed by choosing the following positions for 3 open strings as
 \beqa
 x_{1}=0 ,\qquad x_{2}=1,\qquad x_{3}\rightarrow \infty.
\label{x123int34}\eeqa


\section*{Acknowledgements}

I am indebted to R. Russo, A. Tseytlin, W. Siegel, A. Sen and N. Arkani-Hamed for helpful discussions. I would like to
thank  M. Bianchi, C. Hull, W. Lerche, L. Alvarez-Gaume, M. Douglas, P. Vanhove, H. Steinacker, I. Bena, 
J. Polchinski and P. Sulkowski  for various valuable discussions. I would also like to thank CERN for the hospitality. Part of 
this work was done during 
 the author's 2nd post doctoral position at Queen Mary University of London (QMUL) and he warmly thanks  A. Brandhuber, D. Young, N. Lambert and B. Stefanski for enjoyable discussions. This work was supported by  ERC Starting Grant no. 335739
'Quantum fields and knot homologies', funded by the European Research Council.


\end{document}